\documentclass[sigconf]{acmart}

\AtBeginDocument{%
  }

\usepackage{booktabs} 
\usepackage{amsmath}
\usepackage{graphicx}
\usepackage{setspace}
\usepackage{xcolor}
\usepackage{program}
\usepackage{multirow}
\usepackage{xspace}
\usepackage{pifont}
\usepackage{dirtytalk}
\usepackage{epigraph}
\usepackage{wrapfig}

 \epigraphsize{\small}
\copyrightyear{2023}
\acmYear{2023}
\setcopyright{rightsretained}
\acmConference[CHI EA '23]{Extended Abstracts of the 2023 CHI Conference on Human Factors in Computing Systems}{April 23--28, 2023}{Hamburg, Germany}
\acmBooktitle{Extended Abstracts of the 2023 CHI Conference on Human Factors in Computing Systems (CHI EA '23), April 23--28, 2023, Hamburg, Germany}\acmDOI{10.1145/3544549.3585807}
\acmISBN{978-1-4503-9422-2/23/04}

\DeclareRobustCommand{\company}{Megagon Labs\xspace}
\DeclareRobustCommand{\system}{{\sc Magneton}\xspace}

\newcommand{\stitle}[1]{\vspace{0.2em}\noindent\textbf{#1}}
\newcommand{\hide}[1]{}
\newcommand{\eg}{{\itshape e.g.}, }
\newcommand{\ie}{{\itshape i.e.}, }
\newcommand{\code}[1]{\texttt{\small #1}}

\begin{document}

\title{Towards Transparent, Reusable, and Customizable Data Science in Computational Notebooks}

\author{Frederick Choi}
\authornote{Work done during internship at Megagon Labs.}
\email{fc20@illinois.edu}
\affiliation{%
  \institution{University of Illinois at Urbana-Champaign}
  \city{Urbana}
  \state{Illinois}
  \country{USA}
}
\author{Sajjadur Rahman}
\email{sajjadur@megagon.ai}
\affiliation{%
  \institution{Megagon Labs}
  \city{Mountain View}
  \state{California}
  \country{USA}
}
\author{Hannah Kim}
\email{hannah@megagon.ai}
\affiliation{%
  \institution{Megagon Labs}
  \city{Mountain View}
  \state{California}
  \country{USA}
}
\author{Dan Zhang}
\email{dan_z@megagon.ai}
\affiliation{%
  \institution{Megagon Labs}
  \city{Mountain View}
  \state{California}
  \country{USA}
}

\begin{abstract}
Data science workflows are human-centered processes involving on-demand programming and analysis. While programmable and interactive interfaces such as widgets embedded within computational notebooks are suitable for these workflows, they lack robust state management capabilities and do not support user-defined customization of the interactive components. The absence of such capabilities hinders workflow reusability and transparency while limiting the scope of exploration of the end-users. In response, we developed \system, a framework for authoring interactive widgets within computational notebooks that enables transparent, reusable, and customizable data science workflows. The framework enhances existing widgets to support fine-grained interaction history management, reusable states, and user-defined customizations. We conducted three case studies in a real-world knowledge graph construction and serving platform to evaluate the effectiveness of these widgets. Based on the observations, we discuss future implications of employing \system widgets for general-purpose data science workflows.

\end{abstract}

\ccsdesc[500]{Human-centered computing~Interactive systems and tools}
\ccsdesc[500]{Information systems}
\ccsdesc[500]{Human-centered computing~Visualization}

\keywords{Literate Programming; Exploratory Programming; Interactive programming; Data Science; Computational notebooks;}

\maketitle

\section{Introduction}
Data science workflows are iterative wherein users often switch between multiple tools, including programming environments (\eg Jupyter Notebooks), visualization tools (\eg Tableau or PowerBI), and  spreadsheets (\eg Excel)~\cite{rahman2020leam,wongsuphasawat2019goals}.
Such back-and-forth switching results in a discontinuous workflow --- users are forced to execute repetitive \emph{glue} tasks manually
to bridge the gap between two systems during each switch~\cite{chattopadhyay2020s}.
The overhead of frequent context switching discourages users from using analysis 
tools and restricts them to working
only in code~\cite{wongsuphasawat2019goals}. 
Alspaugh et al.~\cite{futzing2019moseying} advocated for systems
that combine the expressivity of coding
platforms and the ease of use of visual analysis tools.
Computational notebooks (\eg Jupyter~\cite{kluyver2016jupyter} and Observable~\cite{observable}) embedded with interactive interfaces called \emph{widgets}~\cite{IPyWidgets}
support these objectives.
Data practitioners
employ these interfaces for auditing, exploring, and sharing data insights~\cite{wu2020b2,bauerle2022symphony,kery2020mage, rahman2020leam}. 

\begin{figure*}[!htb] 
  \centering
  \includegraphics[width=\linewidth]{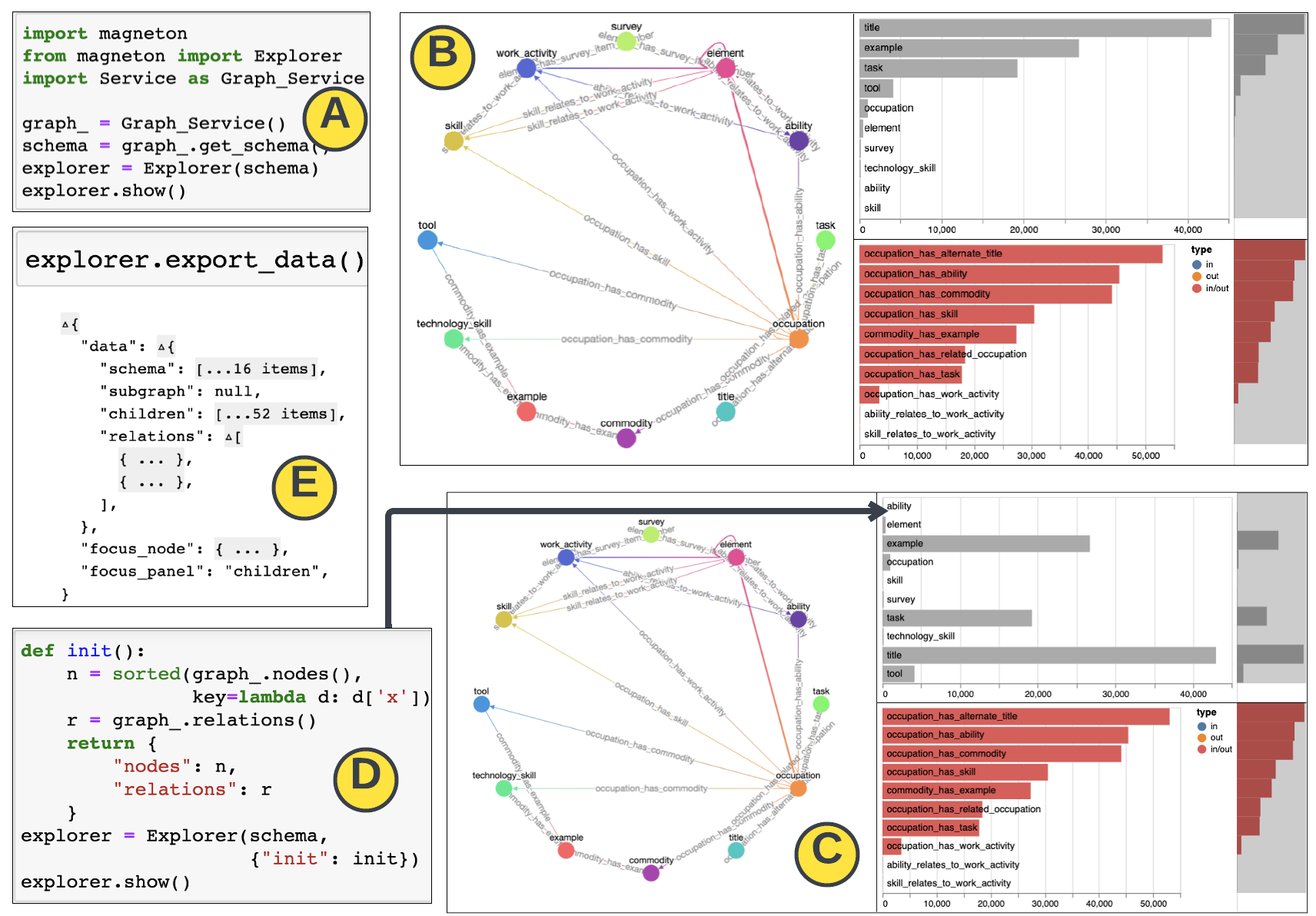}
  \caption{Overview of \system features (counter clock-wise). (A) User instantiates a graph exploration widget from the notebook. (B) A multiple-coordinated view consisting of a graph schema and corresponding node (top) and relation (bottom) distribution components is displayed. (C) A customized widget displaying node distribution in alphabetic order --- (D) 
  user defines an initialization function \code{init()} to customize the sort order and passes it as a callback function during widget initialization. (E) User exports the widget state using the \code{export\_data()} accessor function. }
  \label{fig:teaser} 
  \Description{Overview of the Magneton features in a counter clock-wise fashion. (A) User instantiates a graph exploration widget from the notebook. (B) A multiple-coordinated view consisting of a graph schema and corresponding node (top) and relation (bottom) distribution components is displayed. (C) A customized widget displaying node distribution in alphabetic order --- (D) 
  user defines an intialization function which customizes the sort order and passes it as a callback function during widget initialization. (E) User exports the current widget state using the data export accessor function.}
\end{figure*}


However, gaps remain with respect to transparency, reusability, and customizability of user actions while using these programmable and interactive interfaces. 
Widgets operate as embeddable and lightweight interfaces with interactive components. User actions on the front-end trigger pre-defined operations called \emph{data operations} that update the widget state and re-render components accordingly. The widget state maintains the values of the front-end component properties, \eg frequency distribution corresponding to a bar chart.
Figure~\ref{fig:teaser}B displays such an interface that summarizes a graph database --- the component on the left displays the schema with node and relation types, and the two components on the right shows the corresponding node (top) and relation (bottom) distributions. Clicking a node type in the schema triggers data operations that recompute the corresponding distributions of the bar charts.
Existing widgets such as \emph{ipywidgets}~\cite{IPyWidgets} 
only maintain the most recent state and do not instrument mechanisms to track the state transitions triggered by user interactions.
Therefore, these widgets lack transparency as the history of interactions and their corresponding states are lost and, 
reusability as recovering previous states requires users to execute the interactions from scratch.  
Lack of such state management 
capabilities contribute to loss of knowledge when data science teams share information~\cite{zhang2020data} and limit the reproducibility of workflows~\cite{Kery2018InteractionsFU, kery2019towards36, rahman2022ie}. Furthermore, these interfaces lack the affordances for end-users to customize the built-in data operations. For example, users cannot override the data operation to update the sort order of the bar charts in Figure~\ref{fig:teaser}B from descending order of frequency to alphabetic order of the labels of the bars. However, existing work advocates for such on-demand customizations to ensure more flexibility in exploring diverse objectives~\cite{rahman2022ie, teddy2020chi}. 




To address these gaps, we implement \system\footnote{Similar to Magneton, a robot-like Pok\'emon, our proposed framework stitches together three objectives (Magnemite): transparency, reusability, and customization, to enable robust \emph{programmable and interactive} interfaces in computational notebooks.}, a framework for authoring programmable and interactive interfaces 
equipped with built-in state management and on-demand customization capabilities. These interfaces are developed by extending existing widgets.
Users can author their workflows in the Jupyter Notebook (Figure~\ref{fig:teaser}A) by instantiating task-specific widgets (Figure~\ref{fig:teaser}B). Users can \textbf{customize} an interaction's underlying data operation --- \eg the top bar chart in Figure~\ref{fig:teaser}C --- by writing custom code in the notebook (Figure~\ref{fig:teaser}D). 
Each widget has a built-in history view, enabling users to explore their interaction history and access corresponding states, thereby ensuring \textbf{transparency} and \textbf{reusability} of widget states. Users can export the widget state programmatically as \code{JSON} objects (Figure~\ref{fig:teaser}E). 

We implemented a suite of \system widgets
to support various tasks within an in-house knowledge graph construction and serving platform at \company, such as graph curation and knowledge integration. 
We conducted three preliminary case studies involving these tasks where  
data practitioners at \company used \system within their real-world workflows.
The interactive interfaces embedded within computational notebooks
positively impacted their experiences and helped uncover
interesting insights which remained unnoticed in their regular workflows without \system. Examples include identifying low-quality knowledge acquisition candidates and incorrect knowledge integration recommendations. 
Participants found the on-demand customization feature empowering and the ability to access interactions states helpful in managing \emph{glue} tasks between various steps within their workflows. However, participants also identified several limitations that encourage fundamental research questions related to the learning curve, composing widgets, and balancing automation with user agency. We open-source the \system framework at \url{https://github.com/megagonlabs/magneton}.

\section{Related Work}
\label{sec:related}

Data practitioners often use interactive programming environments
which enable interactive exploration of data~\cite{kery2018story,rahman2022ie}. Computational notebooks are examples
of such environments commonly used by practitioners. Charting libraries such as Altair~\cite{vanderplas2018altair} and Plotly~\cite{plotly} also enable interactive visualization. However, unlike \system these libraries lack affordances for customizing the data operations and do not capture interaction history. 
B2~\cite{wu2020b2}, a dataframe wrapper with chart recommendation capabilities, captures interaction history by transforming UI interactions into dataframe operations and appending those operations in a notebook cell. The intermediate states can be reproduced by uncommenting and executing a python statement which is a cumbersome experience. Moreover, removing the notebook cell erases the interaction history. Therefore, additional version control mechanisms are required to achieve true persistence~\cite{brachmann2020your}. Additionally, the mapping from user interactions to dataframe operations are pre-defined and cannot be customized.
In contrast, \system extends widgets in computational notebooks to enable persistent interaction history and on-demand customization. While B2 is limited to tabular dataframes, \system may support other data domains as users can plug in any functions to overwrite the underlying data operation corresponding to an interaction.
\emph{mage}~\cite{kery2020mage} is another tool similar to B2 that translates interactions on interactive components within widgets to code, \eg dataframe operations, and exhibits similar limitations. Other interactive programming environments are not limited to the notebook paradigm. For example, interactions with visualizations in GUESS~\cite{adar2006guess} and Leam~\cite{rahman2020leam, griggs2021towards} are captured in a Python environment. While visualizations in these tools are programmable via declarative commands, they do not track and persist the user's interaction history, thereby impeding transparency and reusability
of user actions. Variolite~\cite{kery2017variolite} enables users to explore their previous interactions with code only. However, these bespoke tools require users to transition to and learn a new platform. 

Frameworks such as Panel~\cite{panel},
Plotly Dash~\cite{plotlydash}, and Symphony~\cite{bauerle2022symphony} 
use independent components to create visualizations that can be used in both Jupyter notebooks and standalone
websites. However, Plotly Dash does not easily extend to custom visualizations, unlike Panel and Symphony.
\system also supports a wider range of visualization libraries. 
However, these frameworks lack support for on-demand customization of underlying data operations and do not track interaction history, both of which are supported as built-in features by \system widgets.
 Streamlit~\cite{streamlit} is another platform for generating interactive web dashboards using Python script. However, the platform prioritizes web applications rather than exploratory data science and does not focus on objectives such as transparency, reusability, and customization. Moreover, a significant learning curve is associated with learning a new platform.

\section{\system Framework Design}
\label{sec:system}
We now explain how \system helps author widgets that support transparent, reusable, and customizable user actions. 


\subsection{Widget Frameworks: Design and Limitations}
Widgets are interactive elements, \eg sliders, text boxes, buttons, that have representations both in the kernel, \ie where code is executed, and the front-end, \ie the notebook web interface. However, recent frameworks for authoring widgets~\cite{idomjp} also enable integration of interactive dashboards in the front-end~\cite{wu2020b2,bauerle2022symphony, zhang2023meganno}.  

 \begin{figure}[!htb] 
 \centering
  \includegraphics[width=0.8\linewidth]{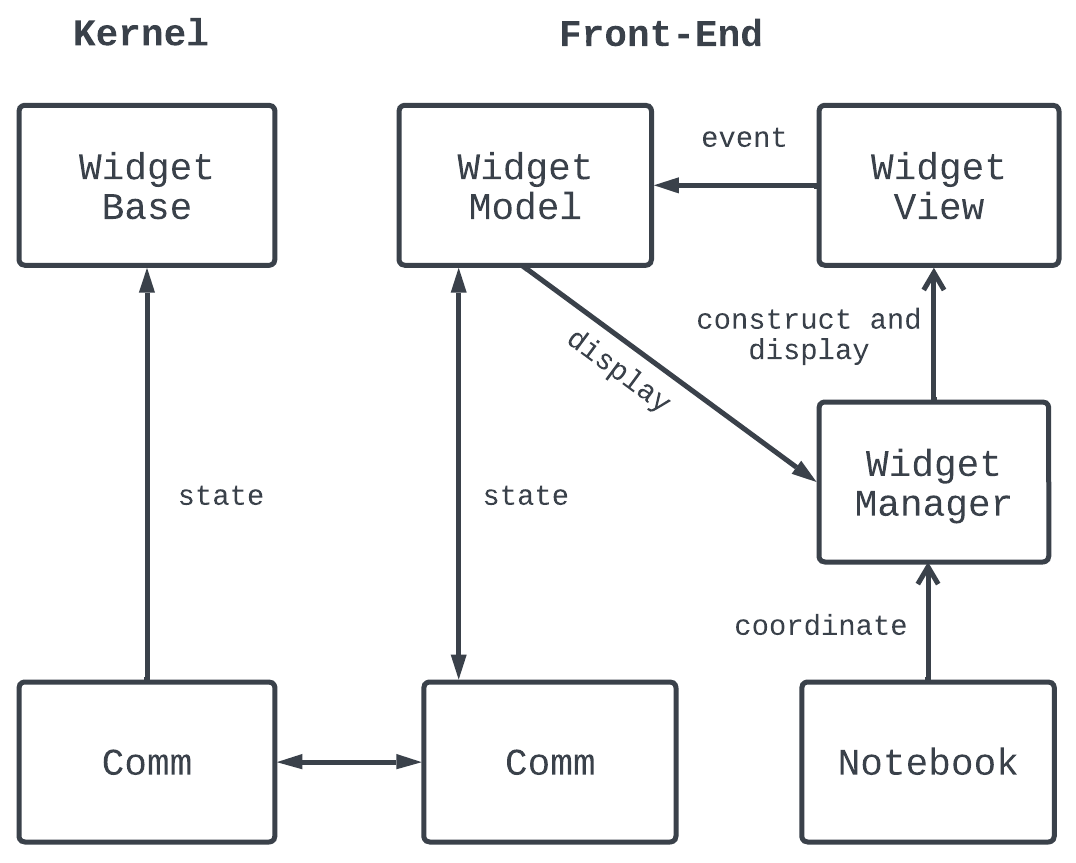}
  \caption{Design of basic, \ie traditional widgets.}
  \label{fig:base_widget} 
  \Description{The basic widget design.}
\end{figure}

 As shown in Figure~\ref{fig:base_widget}, Widgets (\eg \emph{ipywidgets}~\cite{IPyWidgets}) maintain their state both at the back-end kernel (called \emph{Widget Base}) and the front-end (called \emph{Widget Model}.) The Widget Base and Widget Model remain in-sync via the communication API called \emph{Comm}. 
However, only the most recent state is maintained, making the widgets essentially \emph{memoryless}. The \emph{Widget Manager} coordinates the display of the widget in the front-end \emph{Widget View}. The Widget View is a container for rendering interactive components using front-end libraries and web frameworks. The Widget View only registers low-level event listeners corresponding to user interactions on the components. 
 For example, a \emph{selection} interaction on the graph node in Figure~\ref{fig:teaser}B that updates the bar charts is registered as an \emph{onClick} listener.
 Therefore, these widgets are \emph{agnostic} of the user's high-level interaction types and additional context, such as where the interaction happened and which components were updated. The \emph{memoryless} and \emph{interaction agnostic} nature of widgets prevent tracking of the user's interaction history and the corresponding widget states.
Moreover, such a design primarily serves to parameterize data operations in the kernel using front-end events --- a widget state variable (\eg current node identifier) impacted by a low-level event (\eg \emph{onClick}) serves as an input parameter to a data operation (\eg distribution computation). Any change in the widget variable triggers a recomputation of the data operation. In the notebook, the users can programmatically access and update the parameters of the data operations. However, the data operations in the kernel, designed by widget developers, are neither accessible nor customizable from the front-end. The lack of affordances to override data operations limit 
the end-user's capability to customize the widgets designed by the developers. We describe enhancement of existing widgets with such features next.

\subsection{Towards Persistent, Interaction-Aware, and Customizable Widgets}

\begin{figure}[!htb] 
  \centering
  \includegraphics[width=\linewidth]{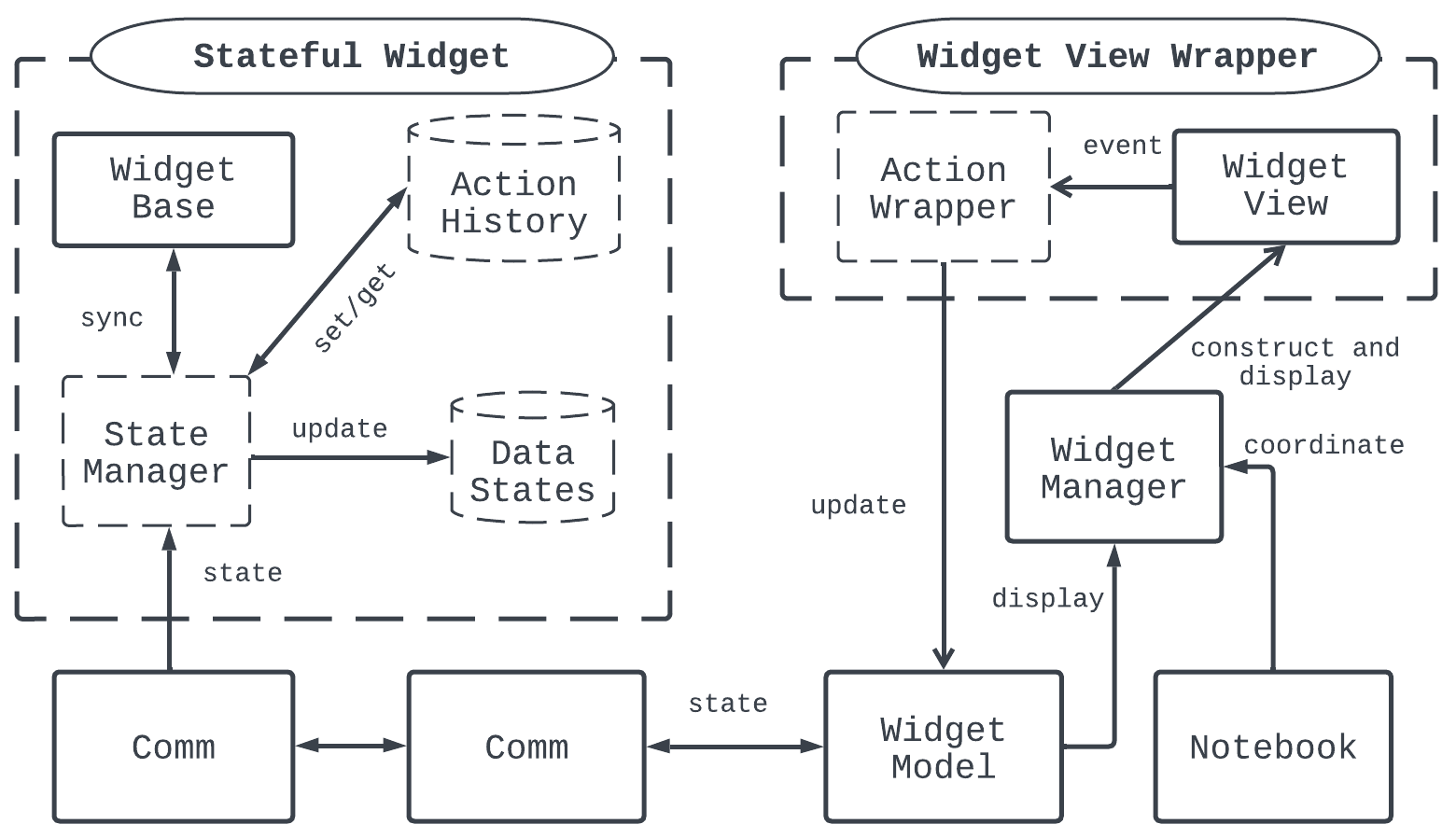}
  \caption{Design of \system widgets. The dashed (``- -'') elements, \ie the stateful widget and widget view wrappers, are introduced by \system.}
  \label{fig:stateful_widget} 
  \Description{The stateful widget design.}
\end{figure}

\begin{figure*}[!htb] 
  \centering
  \includegraphics[width=\linewidth]{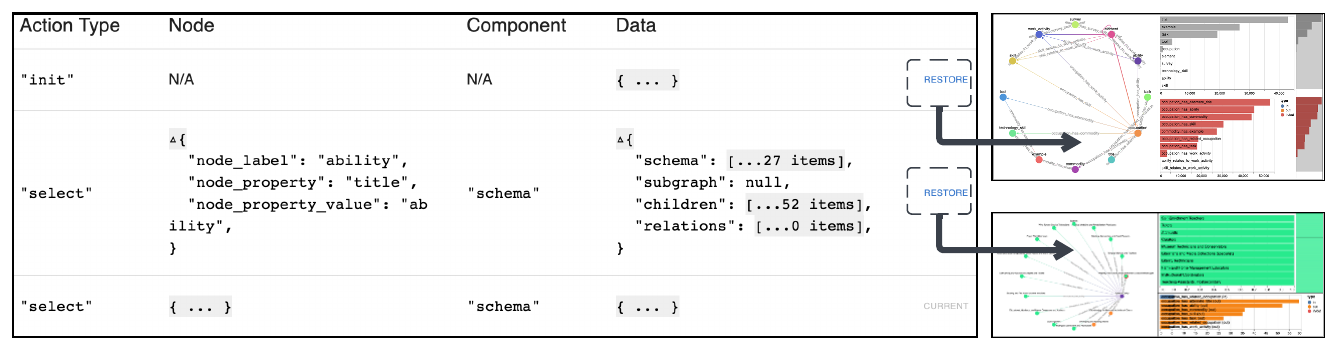}
  \caption{The history view of a widget (\code{widget.history.show()}). Clicking the \emph{Restore} button loads previous state visualizations. }
  \label{fig:history} 
  \Description{The history view of a widget. Clicking the Restore button loads the previous states and their visualizations.}
\end{figure*}

We create a persistent and interaction-aware widget called \emph{stateful widget} by extending the Widget Base with state and interaction history management capabilities (see Figure~\ref{fig:stateful_widget}.) Within a stateful widget, the state manager maintains each state updates corresponding to user interactions within a list called \emph{Data States}. The state manager registers the following in the \emph{action history}: (a) context of each event (\eg the front-end interaction type and the component and element where interaction occurred) and (b) the corresponding state identifier in Data States. Since the default Widget View only registers low-level events, we create a Widget View Wrapper that records each event's context as an action via an Action Wrapper. The action wrapper dispatches an action consisting of the event context mentioned earlier via the Comm API. Users can view the interaction history in a separate notebook cell which shows the details of an interaction and the corresponding data state as shown in Figure~\ref{fig:history}, thereby ensuring transparency. The history view is synchronized with the corresponding widget. Therefore, users can leverage the history to load previous states in the Widget View using the \emph{Restore} button. Moreover, users can also access the widget state as a \code{JSON} object using a declarative command as shown in Figure~\ref{fig:teaser}E, thereby ensuring reusability. Such a design also enables users to employ visualizations as a medium for capturing 
and exporting ``actions interactively
performed in the component''~\cite{batch2017interactive} --- 
the outcomes of these interactions are often utilized in 
subsequent steps of a data science workflow~\cite{rahman2022ie}.

Since data operations in the kernel correspond to user interactions in the front-end component, we introduce the concept of \emph{shared actions}.
Shared actions are data operations that end-users can override from the notebook. The operation definitions are essentially shared between the kernel and front-end. In the \system framework, developers can define a data operation to be shared. 
 For example, a shared data operation may return a distribution sorted by descending order of frequency. However, the user may prefer viewing the distribution in the alphabetic order of labels. As shown in Figure~\ref{fig:teaser}C and~\ref{fig:teaser}D, a user-defined function (UDF) written in the notebook --- which reflects the updated sort order --- is mapped to these the actions during widget instantiation time. 
In the kernel, the state manager parses the UDFs using custom serializers and overrides the data operation corresponding to the shared action. 
Such a design expands the ``events parameterizing code'' paradigm of widgets to ``operations parameterizing code'' and offers more flexible customization capabilities --- users can keep updating the shared actions to explore different objectives by modifying the function defined in the notebook.
Note that developers may implement data operations such as schema generation and distributions computation using standalone libraries or from scratch. In the case studies described in Section~\ref{sec:study}, we used an in-house graph query library, which was published as a Python package. 

\stitle{Components in \system Widget View.} 
We use the React web framework~\cite{react} to develop the front-end components and the IDOM-Jupyter package~\cite{idomjp} for component rendering in the Widget View. 
The components are TypeScript~\cite{typescript} modules that enable the rendering of a wide range web-based visualization libraries. For example, we used a custom graph visualization library to render the schema graph~\cite{franz2016cytoscape}, Vega-lite~\cite{satyanarayan2016vega} to render the bar charts, and a JavaScript library to render tables. 
As TypeScript supports static typing, developers can define application-specific data types and use those across the modules. 
Therefore, using TypeScript ensures a tighter integration between the Widget Base in the kernel and the Widget Model in the front-end. 
Moreover, when customizing data operations defined as shared actions, the pre-defined types provide hints to the user about the expected return type of the customized function. Each of the components rendered in the Widget View is fully interactive. These interactions, derived from existing visualization research~\cite{yi2007toward, amar2005low} are reactively synchronized across \system components, enabling multiple-coordinated visualizations (\eg Figure~\ref{fig:teaser}B.) 

\section{Case Study}
\label{sec:study} 
We conducted three case studies, each involving one participant (one female and two male), related to tasks such as knowledge curation and integration within an in-house knowledge graph construction and serving platform at \company. 
Each study lasted for about an hour. In the first phase of the study, we first provided a brief overview of the general capabilities of
the widgets. We then provided each participant with a Jupyter notebook with relevant \system widgets already imported. 
In the next phase, the participants imported their data and 
then employed the \system widgets
to accomplish their tasks. 
Finally, we asked the participants about their experience using \system widgets. 
Note that the case studies involved \company's proprietary knowledge graph in HR domain and datasets, \eg a job description corpus. Therefore, for the screenshots in the paper, we used knowledge graphs constructed from an open-source resource called O*NET~\cite{peterson1999occupational}. However, the information conveyed in the screenshots reflects the original work setting and experience of the participants. 

\stitle{Case Study Details.} Two case studies involved graph exploration for curating knowledge graphs and identifying opportunities for graph expansion. For graph exploration, the participants employed an \emph{Explorer} widget (Figure~\ref{fig:teaser}A). To explore the graph, the participants performed ($P1$ and $P2$) various interactions on the widget components. Examples include panning and zooming to understand the graph schema and selection to view node and relation distribution which provides additional details helpful for uncovering inconsistencies that require further curation. 
In the final case study, the participant ($P3$) performed a knowledge integration task where they focused on verifying alignment candidates extracted from a text corpus and assigning merging decisions, such as insert, ignore, and defer, by exploring candidate entities in the in-house knowledge graph. We provided an \emph{alignment-verification} widget for the verification task (see Figure~\ref{fig:alignment-verification}). The widget contains an interactive table component with alignment candidates from the text corpus and the graph in two columns and a selection menu of alignment decisions in another column. The widget also displays the context of alignment candidates --- a table component showing descriptions of an entity extracted from the text and a graph component displaying the sub-graph of the corresponding graph node. 

\subsection{Study Observations}

\begin{figure*}[!htb] 
  \centering
  \includegraphics[width=0.8\linewidth]{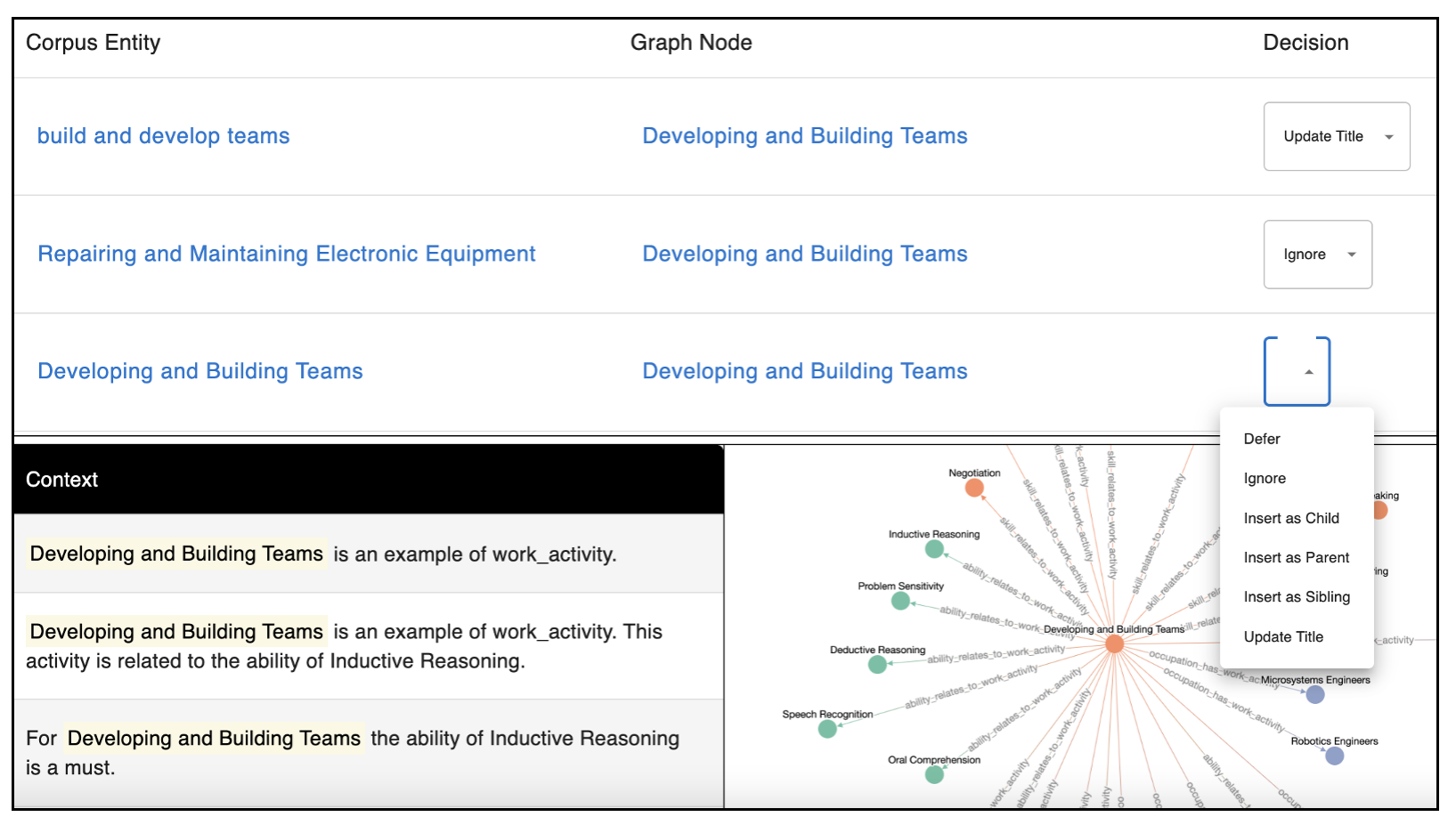}
  \caption{Alignment verification widget.}
  \label{fig:alignment-verification} 
  \Description{Components in the alignment verification widget.}
\end{figure*}

\stitle{\system Widget Usage.} 
The interactive widget-based setup enabled the participant to stitch together different tasks in the same ecosystem, such as writing code and interactively exploring data. In their previous setting, all participants had to switch among multiple tools, such as spreadsheets, notebooks or IDEs, and Neo4j graph browser~\cite{miller2013graph}, which was cumbersome. $P3$ commented --- ``$\ldots$ writing code and visualization; this interactive and graphical feature is much better.'' $P3$ appreciated the ability to view the alignment candidates in the context of the corpus and graph (as shown by the bottom two components in Figure~\ref{fig:alignment-verification}) and make decisions more confidently. Participants ($P1$ and $P2$) were able to identify low-quality long-tail nodes and relations in the graph using the exploration widget. The multiple-coordinated views helped explore node and relation distributions --- low-frequency nodes in the graph were good candidates for further expansion. 
$P1$ commented --- ``I like the feature that you can filter the graph by selecting a node and a corresponding incoming or outgoing relation.'' 

\stitle{Impact of \system Features.} One participant ($P1$) greatly appreciated the general features of \system widgets, such as the history view, which enabled them to explore their interaction history to revisit specific states in the widget. $P1$ mentioned --- ``I wish that it (history view) were integrated within Neo4j browser''. Moreover, the data accessor feature helped participants access the data underlying a given state. $P1$ also commented that the insights obtained through the history view and data obtained through the accessor functions could be shared with other team members during team meetings.
$P2$ utilized the shared action feature to override the data operation underlying the distribution component. They employed a sort wrapper to customize the order of the bars alphabetically to access nodes that the participant was interested in quickly. For example, the participant was interested in a node whose title started with ``c'' and could not scroll and locate the node among approximately 1000 bars in the distribution component. The feature helped improve the discoverability of desired information in the presence of too many bars.
$P3$ also used the shared action feature to declutter the ``sub-graph'' component visualization within the verification widget: instead of showing the entire node neighborhood, which may seem cluttered, the participant used a filtering function --- as a wrapper of the underlying sub-graph computation data operation --- to only display the node, and it's parent. The participant characterized the experience as \emph{transient customizability}: ``When I am exploring, I am not attached to one objective. Right now, I am looking at only the parent (and the node), but later maybe I want to also view siblings $\ldots$ so it can be ephemeral. I like the transient customizability.'' 

\stitle{Limitations.} The participants reported several usability issues with the \system widgets.
$P3$ suggested adding interactions such as filtering and grouping to enable more effective exploration of the interaction states in the history view. $P2$ requested mechanisms to avoid scrolling through the entire interaction history, ``it would be helpful to add a bookmark feature similar to web browsers.''  
While the participants utilized the shared action feature to further customize data operations in widgets, they requested adding more features out-of-the-box to minimize such customization requirements wherever possible. $P2$ commented --- ``the shared action feature is helpful, but a sort could have been avoided had there been a search feature in the bar chart component.''
Moreover, to help design UDFs for overriding shared actions, the participants ($P2$ and $P3$) also requested more clarification in the form of ``standardized documentation'' to explain the components of the widget, their data types, and the supported interactions.  

\section{Discussion}
\label{sec:discuss}

\stitle{Study limitations.} We acknowledge that the scope of our evaluation is limited, specifically targeting the graph domain in the broader landscape of data science workflows while focusing on an industrial setting. Moreover, the study participants were employed at a single company. 
While the observed practices
exist widely in industry and academia, the choice of participants inevitably impacted the generalizability of the findings due to organizational norms, policies, and infrastructures. Additional studies could explore \system's usage benefits and limitations in diverse settings.

\stitle{Documentation and usage.} Even so, the observed usage benefits and limitations of case studies highlight several opportunities for improvement. 
While participants appreciated the freedom of customizing operations that helped accommodate ephemeral and ever-evolving objectives within their exploratory analysis, there were suggestions for adding more features by default. This tension indicates the inherent trade-off between user empowerment and user-friendliness. We are conducting participatory design studies within \company to identify features that could be added as built-in functionalities of the widgets. The study also aims to identify additional components for supporting new workflows within the graph construction and serving platform. To improve the comprehension of a widget's features, we plan to implement a \emph{describe} method to explain component definitions, \ie their data types, supported interactions and their corresponding data operations, and the scope of the shared actions. 

\stitle{Authoring strategies for widget developers.} One of the key strengths of the \system framework is the capability to compose task-oriented widgets by combining components. Therefore, a possible research direction can be to investigate various component authoring strategies for widget developers --- for example, via declarative specification (\eg by leveraging the grammar of interactive graphics~\cite{satyanarayan2016vega}) or using direct manipulation-based (\eg Lyra~\cite{satyanarayan2014lyra}).
Graphileon~\cite{graphileon} is a graph-driven dashboard development environment that uses a graph database to store user-interface components (\eg Networks, Tables, Forms) in nodes. Events modeled as relations define the interactions and data flows between the UI components. Future research may explore such view composition strategies for \system widgets.

\stitle{Utilizing interaction provenance.} As highlighted in the case studies, interaction on the interface, even with specific tasks and goals, may lead to many events being recorded. As a result, the history view may become challenging to use due to perceptual scalability limitations~\cite{liu2013immens, bendre2019faster, rahman2021noah}. Future versions of \system can enhance the history view to support typical exploratory data analysis operations such as search and filter. Moreover, these widgets capture the context (where and when an interaction occurred) and scope (specific data domain, \eg graph) of interactions. Managing such interaction provenance may have additional benefits for intelligent agents that utilize interaction history.
For instance, Solas~\cite{epperson2022leveraging} may leverage rich interaction history across sessions to recommend visualizations or subsequent interactions. Future work may explore ways to complement user-driven exploration with more prescriptive guidance besides expanding on existing research on managing notebook provenance~\cite{brachmann2020your}.    

\stitle{Towards scalable data science.} The interaction-aware \system widgets can serve as the presentation layer of provenance-preserving end-to-end systems for data science~\cite{rahman2022ie, rahman2023mhcai}. However, for large-scale systems, the latency of rendering widgets remains a bottleneck. Recent work on enhancing the scalability of Vega visualization generation introduced automatic
server-side scaling via partitioning strategies~\cite{kruchten2022vegafusion}. Similar strategies can be adopted by \system. Other approaches that may be employed to redesign widgets for scale include applying classical database optimization techniques such as caching, pre-fetching, indexing, materialization, and incremental view maintenance on the server side. 

\stitle{The role of collaboration in widget design.} Since notebooks can be collaborative, \system widgets may need to accommodate shared workflows. Collaboration introduces new challenges, such as enforcing access control mechanisms, characterizing the role of users, and instrumenting conflict resolution techniques~\cite{rahman2020mixtape}, all of which indicate the possibility of interesting future research. Within the collaborative setting, another dimension is the plasticity of interfaces~\cite{passi2018trust} --- 
the perceived value of an interface may vary across stakeholders. To this end, another extension could be to equip \system widgets with cross-platform capabilities~\cite{bauerle2022symphony, rahman2023mhcai}. 

\section{Conclusion}
\label{sec:conclusion}
We contribute \system, a framework for composing 
interaction history-aware and customizable widgets 
to enable transparent, reusable, and expressive
data science workflows in computational notebooks. 
Case studies conducted in the industry
setting demonstrated how \system widgets
enabled efficient graph analysis workflows 
compared to participants' previous experience
by surfacing inconsistencies and data quality issues. 
The built-in interaction history enabled sharing of insights
and data among various steps within a project.
The shared action features empowered users
to customize the widget, thereby reducing dependency on developers.
However, similar to any new framework, \system
exhibited several limitations related to
incomplete documentation and gaps in widget feature design,
among others. These challenges encourage interesting
future research.

\begin{acks}    
We would like to thank Rafael Li Chen for his assistance in setting up the initial project infrastructure. Our study participants set aside time for interviews and reflection. We're grateful for their contributions to this work.
\end{acks}
\balance

\bibliographystyle{ACM-Reference-Format}
\bibliography{main}


\begin{thebibliography}{46}


\ifx \showCODEN    \undefined \def \showCODEN     #1{\unskip}     \fi
\ifx \showDOI      \undefined \def \showDOI       #1{#1}\fi
\ifx \showISBNx    \undefined \def \showISBNx     #1{\unskip}     \fi
\ifx \showISBNxiii \undefined \def \showISBNxiii  #1{\unskip}     \fi
\ifx \showISSN     \undefined \def \showISSN      #1{\unskip}     \fi
\ifx \showLCCN     \undefined \def \showLCCN      #1{\unskip}     \fi
\ifx \shownote     \undefined \def \shownote      #1{#1}          \fi
\ifx \showarticletitle \undefined \def \showarticletitle #1{#1}   \fi
\ifx \showURL      \undefined \def \showURL       {\relax}        \fi
\providecommand\bibfield[2]{#2}
\providecommand\bibinfo[2]{#2}
\providecommand\natexlab[1]{#1}
\providecommand\showeprint[2][]{arXiv:#2}

\bibitem[Adar(2006)]%
        {adar2006guess}
\bibfield{author}{\bibinfo{person}{Eytan Adar}.}
  \bibinfo{year}{2006}\natexlab{}.
\newblock \showarticletitle{GUESS: a language and interface for graph
  exploration}. In \bibinfo{booktitle}{\emph{Proceedings of the SIGCHI
  conference on Human Factors in computing systems}}. \bibinfo{publisher}{ACM},
  \bibinfo{address}{New York, NY, USA}, \bibinfo{pages}{791--800}.
\newblock


\bibitem[Alspaugh et~al\mbox{.}(2019)]%
        {futzing2019moseying}
\bibfield{author}{\bibinfo{person}{Sara Alspaugh}, \bibinfo{person}{Nava
  Zokaei}, \bibinfo{person}{Andrea Liu}, \bibinfo{person}{Cindy Jin}, {and}
  \bibinfo{person}{Marti~A. Hearst}.} \bibinfo{year}{2019}\natexlab{}.
\newblock \showarticletitle{Futzing and Moseying: Interviews with Professional
  Data Analysts on Exploration Practices}.
\newblock \bibinfo{journal}{\emph{IEEE Transactions on Visualization and
  Computer Graphics}} \bibinfo{volume}{25}, \bibinfo{number}{1}
  (\bibinfo{year}{2019}), \bibinfo{pages}{22--31}.
\newblock
\urldef\tempurl%
\url{https://doi.org/10.1109/TVCG.2018.2865040}
\showDOI{\tempurl}


\bibitem[Amar et~al\mbox{.}(2005)]%
        {amar2005low}
\bibfield{author}{\bibinfo{person}{Robert Amar}, \bibinfo{person}{James Eagan},
  {and} \bibinfo{person}{John Stasko}.} \bibinfo{year}{2005}\natexlab{}.
\newblock \showarticletitle{Low-level components of analytic activity in
  information visualization}. In \bibinfo{booktitle}{\emph{IEEE Symposium on
  Information Visualization, 2005. INFOVIS 2005.}} \bibinfo{publisher}{IEEE},
  \bibinfo{address}{New York, NY, USA}, \bibinfo{pages}{111--117}.
\newblock


\bibitem[Batch and Elmqvist(2017)]%
        {batch2017interactive}
\bibfield{author}{\bibinfo{person}{Andrea Batch} {and} \bibinfo{person}{Niklas
  Elmqvist}.} \bibinfo{year}{2017}\natexlab{}.
\newblock \showarticletitle{The interactive visualization gap in initial
  exploratory data analysis}.
\newblock \bibinfo{journal}{\emph{IEEE transactions on visualization and
  computer graphics}} \bibinfo{volume}{24}, \bibinfo{number}{1}
  (\bibinfo{year}{2017}), \bibinfo{pages}{278--287}.
\newblock


\bibitem[B{\"a}uerle et~al\mbox{.}(2022)]%
        {bauerle2022symphony}
\bibfield{author}{\bibinfo{person}{Alex B{\"a}uerle},
  \bibinfo{person}{{\'A}ngel~Alexander Cabrera}, \bibinfo{person}{Fred Hohman},
  \bibinfo{person}{Megan Maher}, \bibinfo{person}{David Koski},
  \bibinfo{person}{Xavier Suau}, \bibinfo{person}{Titus Barik}, {and}
  \bibinfo{person}{Dominik Moritz}.} \bibinfo{year}{2022}\natexlab{}.
\newblock \showarticletitle{Symphony: Composing Interactive Interfaces for
  Machine Learning}. In \bibinfo{booktitle}{\emph{CHI Conference on Human
  Factors in Computing Systems}}. \bibinfo{publisher}{ACM},
  \bibinfo{address}{New York, NY, USA}, \bibinfo{pages}{1--14}.
\newblock


\bibitem[Bendre et~al\mbox{.}(2019)]%
        {bendre2019faster}
\bibfield{author}{\bibinfo{person}{Mangesh Bendre}, \bibinfo{person}{Tana
  Wattanawaroon}, \bibinfo{person}{Sajjadur Rahman}, \bibinfo{person}{Kelly
  Mack}, \bibinfo{person}{Yuyang Liu}, \bibinfo{person}{Shichu Zhu},
  \bibinfo{person}{Yu Lu}, \bibinfo{person}{Ping-Jing Yang},
  \bibinfo{person}{Xinyan Zhou}, \bibinfo{person}{Kevin Chen-Chuan Chang},
  {et~al\mbox{.}}} \bibinfo{year}{2019}\natexlab{}.
\newblock \showarticletitle{Faster, higher, stronger: Redesigning spreadsheets
  for scale}. In \bibinfo{booktitle}{\emph{2019 IEEE 35th International
  Conference on Data Engineering (ICDE)}}. \bibinfo{publisher}{IEEE},
  \bibinfo{address}{New York, NY, USA}, \bibinfo{pages}{1972--1975}.
\newblock


\bibitem[Brachmann and Spoth(2020)]%
        {brachmann2020your}
\bibfield{author}{\bibinfo{person}{Michael Brachmann} {and}
  \bibinfo{person}{William Spoth}.} \bibinfo{year}{2020}\natexlab{}.
\newblock \showarticletitle{Your notebook is not crumby enough, REPLace it}. In
  \bibinfo{booktitle}{\emph{Conference on Innovative Data Systems Research
  (CIDR)}}. \bibinfo{publisher}{CIDRDB}, \bibinfo{address}{Chaminade,
  California}, \bibinfo{pages}{1--8}.
\newblock


\bibitem[Chattopadhyay et~al\mbox{.}(2020)]%
        {chattopadhyay2020s}
\bibfield{author}{\bibinfo{person}{Souti Chattopadhyay},
  \bibinfo{person}{Ishita Prasad}, \bibinfo{person}{Austin~Z Henley},
  \bibinfo{person}{Anita Sarma}, {and} \bibinfo{person}{Titus Barik}.}
  \bibinfo{year}{2020}\natexlab{}.
\newblock \showarticletitle{What's wrong with computational notebooks? Pain
  points, needs, and design opportunities}. In
  \bibinfo{booktitle}{\emph{Proceedings of the 2020 CHI Conference on Human
  Factors in Computing Systems}}. \bibinfo{publisher}{ACM},
  \bibinfo{address}{New York, NY, USA}, \bibinfo{pages}{1--12}.
\newblock


\bibitem[Epperson et~al\mbox{.}(2022)]%
        {epperson2022leveraging}
\bibfield{author}{\bibinfo{person}{Will Epperson}, \bibinfo{person}{Doris
  Jung-Lin~Lee}, \bibinfo{person}{Leijie Wang}, \bibinfo{person}{Kunal
  Agarwal}, \bibinfo{person}{Aditya~G Parameswaran}, \bibinfo{person}{Dominik
  Moritz}, {and} \bibinfo{person}{Adam Perer}.}
  \bibinfo{year}{2022}\natexlab{}.
\newblock \showarticletitle{Leveraging Analysis History for Improved In Situ
  Visualization Recommendation}.
\newblock \bibinfo{journal}{\emph{Computer Graphics Forum}}
  \bibinfo{volume}{41}, \bibinfo{number}{3} (\bibinfo{year}{2022}),
  \bibinfo{pages}{145--155}.
\newblock


\bibitem[Franz et~al\mbox{.}(2016)]%
        {franz2016cytoscape}
\bibfield{author}{\bibinfo{person}{Max Franz}, \bibinfo{person}{Christian~T
  Lopes}, \bibinfo{person}{Gerardo Huck}, \bibinfo{person}{Yue Dong},
  \bibinfo{person}{Onur Sumer}, {and} \bibinfo{person}{Gary~D Bader}.}
  \bibinfo{year}{2016}\natexlab{}.
\newblock \showarticletitle{Cytoscape. js: a graph theory library for
  visualisation and analysis}.
\newblock \bibinfo{journal}{\emph{Bioinformatics}} \bibinfo{volume}{32},
  \bibinfo{number}{2} (\bibinfo{year}{2016}), \bibinfo{pages}{309--311}.
\newblock


\bibitem[Graphileon(2022)]%
        {graphileon}
\bibfield{author}{\bibinfo{person}{Graphileon}.}
  \bibinfo{year}{2022}\natexlab{}.
\newblock \bibinfo{booktitle}{\emph{Graphileon}}.
\newblock Graphileon.
\newblock
\urldef\tempurl%
\url{https://graphileon.com/}
\showURL{%
Retrieved January 19, 2023 from \tempurl}


\bibitem[Griggs et~al\mbox{.}(2021)]%
        {griggs2021towards}
\bibfield{author}{\bibinfo{person}{Peter Griggs}, \bibinfo{person}{Cagatay
  Demiralp}, {and} \bibinfo{person}{Sajjadur Rahman}.}
  \bibinfo{year}{2021}\natexlab{}.
\newblock \showarticletitle{Towards integrated, interactive, and extensible
  text data analytics with Leam}. In \bibinfo{booktitle}{\emph{Proceedings of
  the Second Workshop on Data Science with Human in the Loop: Language
  Advances}}. \bibinfo{publisher}{Association for Computational Linguistics},
  \bibinfo{address}{Online}, \bibinfo{pages}{52--58}.
\newblock
\urldef\tempurl%
\url{https://doi.org/10.18653/v1/2021.dash-1.9}
\showDOI{\tempurl}


\bibitem[Holoviz(2022)]%
        {panel}
\bibfield{author}{\bibinfo{person}{Holoviz}.} \bibinfo{year}{2022}\natexlab{}.
\newblock \bibinfo{booktitle}{\emph{Panel}}.
\newblock Panel.
\newblock
\urldef\tempurl%
\url{https://panel.holoviz.org/}
\showURL{%
Retrieved January 19, 2023 from \tempurl}


\bibitem[IDOM(2022)]%
        {idomjp}
\bibfield{author}{\bibinfo{person}{IDOM}.} \bibinfo{year}{2022}\natexlab{}.
\newblock \bibinfo{booktitle}{\emph{IDOM - a declarative Python package for
  building highly interactive user interfaces.}}
\newblock IDOM.
\newblock
\urldef\tempurl%
\url{https://ryanmorshead.com/articles/2021/idom-react-but-its-python/article/}
\showURL{%
Retrieved January 19, 2023 from \tempurl}


\bibitem[Inc.(2022a)]%
        {observable}
\bibfield{author}{\bibinfo{person}{Observable Inc.}}
  \bibinfo{year}{2022}\natexlab{a}.
\newblock \bibinfo{booktitle}{\emph{Observable}}.
\newblock Observable.
\newblock
\urldef\tempurl%
\url{https://observablehq.com/}
\showURL{%
Retrieved January 19, 2023 from \tempurl}


\bibitem[Inc.(2022b)]%
        {streamlit}
\bibfield{author}{\bibinfo{person}{Streamlit Inc.}}
  \bibinfo{year}{2022}\natexlab{b}.
\newblock \bibinfo{booktitle}{\emph{Streamlit}}.
\newblock Streamlit.
\newblock
\urldef\tempurl%
\url{https://streamlit.io/}
\showURL{%
Retrieved January 19, 2023 from \tempurl}


\bibitem[Jupyter(2022)]%
        {IPyWidgets}
\bibfield{author}{\bibinfo{person}{Jupyter}.} \bibinfo{year}{2022}\natexlab{}.
\newblock \bibinfo{booktitle}{\emph{IPyWidgets}}.
\newblock Jupyter.
\newblock
\urldef\tempurl%
\url{https://ipywidgets.readthedocs.io/en/stable/}
\showURL{%
Retrieved January 19, 2023 from \tempurl}


\bibitem[Kery et~al\mbox{.}(2017)]%
        {kery2017variolite}
\bibfield{author}{\bibinfo{person}{Mary~Beth Kery}, \bibinfo{person}{Amber
  Horvath}, {and} \bibinfo{person}{Brad Myers}.}
  \bibinfo{year}{2017}\natexlab{}.
\newblock \showarticletitle{Variolite: Supporting Exploratory Programming by
  Data Scientists}. In \bibinfo{booktitle}{\emph{Proceedings of the 2017 CHI
  Conference on Human Factors in Computing Systems}}. \bibinfo{publisher}{ACM},
  \bibinfo{address}{New York, NY, USA}, \bibinfo{pages}{1265–1276}.
\newblock


\bibitem[Kery et~al\mbox{.}(2019)]%
        {kery2019towards36}
\bibfield{author}{\bibinfo{person}{Mary~Beth Kery}, \bibinfo{person}{Bonnie~E
  John}, \bibinfo{person}{Patrick O'Flaherty}, \bibinfo{person}{Amber Horvath},
  {and} \bibinfo{person}{Brad~A Myers}.} \bibinfo{year}{2019}\natexlab{}.
\newblock \showarticletitle{Towards effective foraging by data scientists to
  find past analysis choices}. In \bibinfo{booktitle}{\emph{Proceedings of the
  2019 CHI Conference on Human Factors in Computing Systems}}.
  \bibinfo{publisher}{ACM}, \bibinfo{address}{New York, NY, USA},
  \bibinfo{pages}{1--13}.
\newblock


\bibitem[Kery and Myers(2018)]%
        {Kery2018InteractionsFU}
\bibfield{author}{\bibinfo{person}{Mary~Beth Kery} {and}
  \bibinfo{person}{Brad~A. Myers}.} \bibinfo{year}{2018}\natexlab{}.
\newblock \showarticletitle{Interactions for Untangling Messy History in a
  Computational Notebook}. In \bibinfo{booktitle}{\emph{2018 IEEE Symposium on
  Visual Languages and Human-Centric Computing (VL/HCC)}}.
  \bibinfo{publisher}{IEEE}, \bibinfo{address}{New York, NY, USA},
  \bibinfo{pages}{147--155}.
\newblock


\bibitem[Kery et~al\mbox{.}(2018)]%
        {kery2018story}
\bibfield{author}{\bibinfo{person}{Mary~Beth Kery}, \bibinfo{person}{Marissa
  Radensky}, \bibinfo{person}{Mahima Arya}, \bibinfo{person}{Bonnie~E John},
  {and} \bibinfo{person}{Brad~A Myers}.} \bibinfo{year}{2018}\natexlab{}.
\newblock \showarticletitle{The story in the notebook: Exploratory data science
  using a literate programming tool}. In \bibinfo{booktitle}{\emph{Proceedings
  of the 2018 CHI Conference on Human Factors in Computing Systems}}.
  \bibinfo{publisher}{ACM}, \bibinfo{address}{New York, NY, USA},
  \bibinfo{pages}{1--11}.
\newblock


\bibitem[Kery et~al\mbox{.}(2020)]%
        {kery2020mage}
\bibfield{author}{\bibinfo{person}{Mary~Beth Kery}, \bibinfo{person}{Donghao
  Ren}, \bibinfo{person}{Fred Hohman}, \bibinfo{person}{Dominik Moritz},
  \bibinfo{person}{Kanit Wongsuphasawat}, {and} \bibinfo{person}{Kayur Patel}.}
  \bibinfo{year}{2020}\natexlab{}.
\newblock \showarticletitle{mage: Fluid moves between code and graphical work
  in computational notebooks}. In \bibinfo{booktitle}{\emph{Proceedings of the
  33rd Annual ACM Symposium on User Interface Software and Technology}}.
  \bibinfo{publisher}{ACM}, \bibinfo{address}{New York, NY, USA},
  \bibinfo{pages}{140--151}.
\newblock


\bibitem[Kluyver et~al\mbox{.}(2016)]%
        {kluyver2016jupyter}
\bibfield{author}{\bibinfo{person}{Thomas Kluyver}, \bibinfo{person}{Benjamin
  Ragan-Kelley}, \bibinfo{person}{Fernando P{\'e}rez}, \bibinfo{person}{Brian~E
  Granger}, \bibinfo{person}{Matthias Bussonnier}, \bibinfo{person}{Jonathan
  Frederic}, \bibinfo{person}{Kyle Kelley}, \bibinfo{person}{Jessica~B
  Hamrick}, \bibinfo{person}{Jason Grout}, \bibinfo{person}{Sylvain Corlay},
  {et~al\mbox{.}}} \bibinfo{year}{2016}\natexlab{}.
\newblock \bibinfo{booktitle}{\emph{Jupyter Notebooks-a publishing format for
  reproducible computational workflows.}} Vol.~\bibinfo{volume}{2016}.
\newblock \bibinfo{publisher}{IOS Press}, \bibinfo{address}{Amsterdam,
  Netherlands}.
\newblock


\bibitem[Kontosh(1999)]%
        {peterson1999occupational}
\bibfield{author}{\bibinfo{person}{Larry~G Kontosh}.}
  \bibinfo{year}{1999}\natexlab{}.
\newblock \showarticletitle{An occupational information system for the 21st
  century: The development of the O* NET}.
\newblock \bibinfo{journal}{\emph{Journal of Applied Rehabilitation
  Counseling}} \bibinfo{volume}{30}, \bibinfo{number}{2}
  (\bibinfo{year}{1999}), \bibinfo{pages}{43}.
\newblock


\bibitem[Kruchten et~al\mbox{.}(2022)]%
        {kruchten2022vegafusion}
\bibfield{author}{\bibinfo{person}{Nicolas Kruchten}, \bibinfo{person}{Jon
  Mease}, {and} \bibinfo{person}{Dominik Moritz}.}
  \bibinfo{year}{2022}\natexlab{}.
\newblock \showarticletitle{VegaFusion: Automatic Server-Side Scaling for
  Interactive Vega Visualizations}. In \bibinfo{booktitle}{\emph{2022 IEEE
  Visualization and Visual Analytics (VIS)}}. \bibinfo{publisher}{IEEE},
  \bibinfo{address}{New York, NY, USA}, \bibinfo{pages}{11--15}.
\newblock


\bibitem[Liu et~al\mbox{.}(2013)]%
        {liu2013immens}
\bibfield{author}{\bibinfo{person}{Zhicheng Liu}, \bibinfo{person}{Biye Jiang},
  {and} \bibinfo{person}{Jeffrey Heer}.} \bibinfo{year}{2013}\natexlab{}.
\newblock \showarticletitle{imMens: Real-time visual querying of big data}. In
  \bibinfo{booktitle}{\emph{Computer Graphics Forum}}.
  \bibinfo{publisher}{Wiley Online Library}, \bibinfo{address}{New York, NY,
  USA}, \bibinfo{pages}{421--430}.
\newblock


\bibitem[Miller(2013)]%
        {miller2013graph}
\bibfield{author}{\bibinfo{person}{Justin~J Miller}.}
  \bibinfo{year}{2013}\natexlab{}.
\newblock \showarticletitle{Graph database applications and concepts with
  Neo4j}. In \bibinfo{booktitle}{\emph{Proceedings of the southern association
  for information systems conference, Atlanta, GA, USA}}.
  \bibinfo{publisher}{AIS eLibrary}, \bibinfo{address}{WWW},
  \bibinfo{pages}{1--7}.
\newblock


\bibitem[Passi and Jackson(2018)]%
        {passi2018trust}
\bibfield{author}{\bibinfo{person}{Samir Passi} {and} \bibinfo{person}{Steven~J
  Jackson}.} \bibinfo{year}{2018}\natexlab{}.
\newblock \showarticletitle{Trust in data science: Collaboration, translation,
  and accountability in corporate data science projects}.
\newblock \bibinfo{journal}{\emph{Proceedings of the ACM on Human-Computer
  Interaction}} \bibinfo{volume}{2}, \bibinfo{number}{CSCW}
  (\bibinfo{year}{2018}), \bibinfo{pages}{1--28}.
\newblock


\bibitem[Plotly(2022a)]%
        {plotlydash}
\bibfield{author}{\bibinfo{person}{Plotly}.} \bibinfo{year}{2022}\natexlab{a}.
\newblock \bibinfo{booktitle}{\emph{Dash}}.
\newblock Plotly.
\newblock
\urldef\tempurl%
\url{https://plotly.com/dash/}
\showURL{%
Retrieved January 19, 2023 from \tempurl}


\bibitem[Plotly(2022b)]%
        {plotly}
\bibfield{author}{\bibinfo{person}{Plotly}.} \bibinfo{year}{2022}\natexlab{b}.
\newblock \bibinfo{booktitle}{\emph{Low-code Data Apps}}.
\newblock Plotly.
\newblock
\urldef\tempurl%
\url{https://plotly.com/}
\showURL{%
Retrieved January 19, 2023 from \tempurl}


\bibitem[Rahman et~al\mbox{.}(2021a)]%
        {rahman2021noah}
\bibfield{author}{\bibinfo{person}{Sajjadur Rahman}, \bibinfo{person}{Mangesh
  Bendre}, \bibinfo{person}{Yuyang Liu}, \bibinfo{person}{Shichu Zhu},
  \bibinfo{person}{Zhaoyuan Su}, \bibinfo{person}{Karrie Karahalios}, {and}
  \bibinfo{person}{Aditya Parameswaran}.} \bibinfo{year}{2021}\natexlab{a}.
\newblock \showarticletitle{NOAH: Interactive Spreadsheet Exploration with
  Dynamic Hierarchical Overviews}.
\newblock \bibinfo{journal}{\emph{Proceedings of the VLDB Endowment}}
  \bibinfo{volume}{14}, \bibinfo{number}{6} (\bibinfo{year}{2021}),
  \bibinfo{pages}{970--983}.
\newblock


\bibitem[Rahman et~al\mbox{.}(2021b)]%
        {rahman2020leam}
\bibfield{author}{\bibinfo{person}{Sajjadur Rahman}, \bibinfo{person}{Peter
  Griggs}, {and} \bibinfo{person}{{\c{C}}a{\u{g}}atay Demiralp}.}
  \bibinfo{year}{2021}\natexlab{b}.
\newblock \showarticletitle{Leam: An Interactive System for In-situ Visual Text
  Analysis}. In \bibinfo{booktitle}{\emph{CIDR}}. \bibinfo{publisher}{CIDRDB},
  \bibinfo{address}{cidrdb.org}, \bibinfo{pages}{1--7}.
\newblock


\bibitem[Rahman and Kandogan(2022)]%
        {rahman2022ie}
\bibfield{author}{\bibinfo{person}{Sajjadur Rahman} {and} \bibinfo{person}{Eser
  Kandogan}.} \bibinfo{year}{2022}\natexlab{}.
\newblock \showarticletitle{Characterizing Practices, Limitations, and
  Opportunities Related to Text Information Extraction Workflows: A
  Human-in-the-Loop Perspective}. In \bibinfo{booktitle}{\emph{CHI Conference
  on Human Factors in Computing Systems}} (New Orleans, LA, USA)
  \emph{(\bibinfo{series}{CHI '22})}. \bibinfo{publisher}{Association for
  Computing Machinery}, \bibinfo{address}{New York, NY, USA}, Article
  \bibinfo{articleno}{628}, \bibinfo{numpages}{15}~pages.
\newblock
\showISBNx{9781450391573}
\urldef\tempurl%
\url{https://doi.org/10.1145/3491102.3502068}
\showDOI{\tempurl}


\bibitem[Rahman et~al\mbox{.}(2022)]%
        {rahman2023mhcai}
\bibfield{author}{\bibinfo{person}{Sajjadur Rahman}, \bibinfo{person}{Hannah
  Kim}, \bibinfo{person}{Dan Zhang}, \bibinfo{person}{Estevam Hruschka}, {and}
  \bibinfo{person}{Eser Kandogan}.} \bibinfo{year}{2022}\natexlab{}.
\newblock \showarticletitle{Towards Multifaceted Human-Centered AI}.
\newblock \bibinfo{journal}{\emph{ArXiv}}  \bibinfo{volume}{abs/2301.03656}
  (\bibinfo{year}{2022}), \bibinfo{pages}{1--2}.
\newblock


\bibitem[Rahman et~al\mbox{.}(2020)]%
        {rahman2020mixtape}
\bibfield{author}{\bibinfo{person}{Sajjadur Rahman}, \bibinfo{person}{Pao
  Siangliulue}, {and} \bibinfo{person}{Adam Marcus}.}
  \bibinfo{year}{2020}\natexlab{}.
\newblock \showarticletitle{MixTAPE: Mixed-initiative Team Action Plan Creation
  Through Semi-structured Notes, Automatic Task Generation, and Task
  Classification}.
\newblock \bibinfo{journal}{\emph{Proceedings of the ACM on Human-Computer
  Interaction}} \bibinfo{volume}{4}, \bibinfo{number}{CSCW2}
  (\bibinfo{year}{2020}), \bibinfo{pages}{1--26}.
\newblock


\bibitem[ReactJS(2022)]%
        {react}
\bibfield{author}{\bibinfo{person}{ReactJS}.} \bibinfo{year}{2022}\natexlab{}.
\newblock \bibinfo{booktitle}{\emph{React Framework.}}
\newblock ReactJS.
\newblock
\urldef\tempurl%
\url{https://reactjs.org/}
\showURL{%
Retrieved January 19, 2023 from \tempurl}


\bibitem[Satyanarayan and Heer(2014)]%
        {satyanarayan2014lyra}
\bibfield{author}{\bibinfo{person}{Arvind Satyanarayan} {and}
  \bibinfo{person}{Jeffrey Heer}.} \bibinfo{year}{2014}\natexlab{}.
\newblock \showarticletitle{Lyra: An interactive visualization design
  environment}. In \bibinfo{booktitle}{\emph{Computer Graphics Forum}}.
  \bibinfo{publisher}{Wiley Online Library}, \bibinfo{address}{New York, NY,
  USA}, \bibinfo{pages}{351--360}.
\newblock


\bibitem[Satyanarayan et~al\mbox{.}(2016)]%
        {satyanarayan2016vega}
\bibfield{author}{\bibinfo{person}{Arvind Satyanarayan},
  \bibinfo{person}{Dominik Moritz}, \bibinfo{person}{Kanit Wongsuphasawat},
  {and} \bibinfo{person}{Jeffrey Heer}.} \bibinfo{year}{2016}\natexlab{}.
\newblock \showarticletitle{Vega-lite: A grammar of interactive graphics}.
\newblock \bibinfo{journal}{\emph{IEEE transactions on visualization and
  computer graphics}} \bibinfo{volume}{23}, \bibinfo{number}{1}
  (\bibinfo{year}{2016}), \bibinfo{pages}{341--350}.
\newblock


\bibitem[TypeScript(2022)]%
        {typescript}
\bibfield{author}{\bibinfo{person}{TypeScript}.}
  \bibinfo{year}{2022}\natexlab{}.
\newblock \bibinfo{booktitle}{\emph{TypeScript Language.}}
\newblock TypeScript.
\newblock
\urldef\tempurl%
\url{https://www.typescriptlang.org/}
\showURL{%
Retrieved January 19, 2023 from \tempurl}


\bibitem[VanderPlas et~al\mbox{.}(2018)]%
        {vanderplas2018altair}
\bibfield{author}{\bibinfo{person}{Jacob VanderPlas}, \bibinfo{person}{Brian
  Granger}, \bibinfo{person}{Jeffrey Heer}, \bibinfo{person}{Dominik Moritz},
  \bibinfo{person}{Kanit Wongsuphasawat}, \bibinfo{person}{Arvind
  Satyanarayan}, \bibinfo{person}{Eitan Lees}, \bibinfo{person}{Ilia Timofeev},
  \bibinfo{person}{Ben Welsh}, {and} \bibinfo{person}{Scott Sievert}.}
  \bibinfo{year}{2018}\natexlab{}.
\newblock \showarticletitle{Altair: interactive statistical visualizations for
  Python}.
\newblock \bibinfo{journal}{\emph{Journal of open source software}}
  \bibinfo{volume}{3}, \bibinfo{number}{32} (\bibinfo{year}{2018}),
  \bibinfo{pages}{1057}.
\newblock


\bibitem[Wongsuphasawat et~al\mbox{.}(2019)]%
        {wongsuphasawat2019goals}
\bibfield{author}{\bibinfo{person}{Kanit Wongsuphasawat}, \bibinfo{person}{Yang
  Liu}, {and} \bibinfo{person}{Jeffrey Heer}.} \bibinfo{year}{2019}\natexlab{}.
\newblock \showarticletitle{Goals, Process, and Challenges of Exploratory Data
  Analysis: An Interview Study}.
\newblock \bibinfo{journal}{\emph{ArXiv}}  \bibinfo{volume}{abs/1911.00568}
  (\bibinfo{year}{2019}), \bibinfo{pages}{1--10}.
\newblock


\bibitem[Wu et~al\mbox{.}(2020)]%
        {wu2020b2}
\bibfield{author}{\bibinfo{person}{Yifan Wu}, \bibinfo{person}{Joseph~M
  Hellerstein}, {and} \bibinfo{person}{Arvind Satyanarayan}.}
  \bibinfo{year}{2020}\natexlab{}.
\newblock \showarticletitle{B2: Bridging code and interactive visualization in
  computational notebooks}. In \bibinfo{booktitle}{\emph{Proceedings of the
  33rd Annual ACM Symposium on User Interface Software and Technology}}.
  \bibinfo{publisher}{Association for Computing Machinery},
  \bibinfo{address}{New York, NY, USA}, \bibinfo{pages}{152--165}.
\newblock


\bibitem[Yi et~al\mbox{.}(2007)]%
        {yi2007toward}
\bibfield{author}{\bibinfo{person}{Ji~Soo Yi}, \bibinfo{person}{Youn ah Kang},
  \bibinfo{person}{John Stasko}, {and} \bibinfo{person}{Julie~A Jacko}.}
  \bibinfo{year}{2007}\natexlab{}.
\newblock \showarticletitle{Toward a deeper understanding of the role of
  interaction in information visualization}.
\newblock \bibinfo{journal}{\emph{IEEE transactions on visualization and
  computer graphics}} \bibinfo{volume}{13}, \bibinfo{number}{6}
  (\bibinfo{year}{2007}), \bibinfo{pages}{1224--1231}.
\newblock


\bibitem[Zhang et~al\mbox{.}(2020b)]%
        {zhang2020data}
\bibfield{author}{\bibinfo{person}{Amy~X Zhang}, \bibinfo{person}{Michael
  Muller}, {and} \bibinfo{person}{Dakuo Wang}.}
  \bibinfo{year}{2020}\natexlab{b}.
\newblock \showarticletitle{How do data science workers collaborate? roles,
  workflows, and tools}.
\newblock \bibinfo{journal}{\emph{Proceedings of the ACM on Human-Computer
  Interaction}} \bibinfo{volume}{4}, \bibinfo{number}{CSCW1}
  (\bibinfo{year}{2020}), \bibinfo{pages}{1--23}.
\newblock


\bibitem[Zhang et~al\mbox{.}(2022)]%
        {zhang2023meganno}
\bibfield{author}{\bibinfo{person}{Dan Zhang}, \bibinfo{person}{Hannah Kim},
  \bibinfo{person}{Rafael Li~Chen}, \bibinfo{person}{Eser Kandogan}, {and}
  \bibinfo{person}{Estevam Hruschka}.} \bibinfo{year}{2022}\natexlab{}.
\newblock \showarticletitle{{MEGA}nno: Exploratory Labeling for {NLP} in
  Computational Notebooks}. In \bibinfo{booktitle}{\emph{Proceedings of the
  Fourth Workshop on Data Science with Human-in-the-Loop (Language Advances)}}.
  \bibinfo{publisher}{Association for Computational Linguistics},
  \bibinfo{address}{Abu Dhabi, United Arab Emirates (Hybrid)},
  \bibinfo{pages}{1--7}.
\newblock
\urldef\tempurl%
\url{https://aclanthology.org/2022.dash-1.1}
\showURL{%
\tempurl}


\bibitem[Zhang et~al\mbox{.}(2020a)]%
        {teddy2020chi}
\bibfield{author}{\bibinfo{person}{Xiong Zhang}, \bibinfo{person}{Jonathan
  Engel}, \bibinfo{person}{Sara Evensen}, \bibinfo{person}{Yuliang Li},
  \bibinfo{person}{\c{C}a\u{g}atay Demiralp}, {and} \bibinfo{person}{Wang-Chiew
  Tan}.} \bibinfo{year}{2020}\natexlab{a}.
\newblock \showarticletitle{Teddy: A System for Interactive Review Analysis}.
  In \bibinfo{booktitle}{\emph{Proceedings of the 2020 CHI Conference on Human
  Factors in Computing Systems}} (Honolulu, HI, USA)
  \emph{(\bibinfo{series}{CHI '20})}. \bibinfo{publisher}{Association for
  Computing Machinery}, \bibinfo{address}{New York, NY, USA},
  \bibinfo{pages}{1–13}.
\newblock
\showISBNx{9781450367080}
\urldef\tempurl%
\url{https://doi.org/10.1145/3313831.3376235}
\showDOI{\tempurl}


\end{thebibliography}



\end{document}